\documentclass[draftcls,12pt,onecolumn,dvipdfm]{IEEEtran}  

\usepackage{mathrsfs}
\usepackage{amsfonts}
\usepackage{amssymb,epic, graphicx}
\usepackage[cmex10]{amsmath}
\usepackage{array}
\usepackage{mdwmath}
\usepackage{mdwtab}
\usepackage{eqparbox}
\usepackage[tight,footnotesize]{subfigure}
\usepackage{ifpdf}

\ifCLASSINFOpdf

\else

\fi

\usepackage[cmex10]{amsmath}

\newtheorem{definition}{Definition}
\newtheorem{theorem}{Theorem}
\newtheorem{proposition}{Proposition}
\newtheorem{lemma}{Lemma}

\usepackage[square,numbers,sort&compress]{natbib}

\usepackage{ifpdf}

\begin{document}

\title{New Geometrical Spectra of Linear Codes with Applications to Performance Analysis}

\author{\authorblockN{Xiao Ma\authorrefmark{1}, Jia Liu\authorrefmark{1}\authorrefmark{2}, and Qiutao
Zhuang\authorrefmark{1}\\
\authorblockA{\authorrefmark{1}Department of Electronics and Communication
Engineering, Sun Yat-sen University, Guangzhou 510006, GD, China}\\
\authorblockA{\authorrefmark{2}College of Comp. Sci. and Eng., Zhongkai University of Agriculture and Engineering, Guangzhou 510225, GD, China}
} \\
Email: maxiao@mail.sysu.edu.cn, ljia2@mail2.sysu.edu.cn and
zhuangqt@mail2.sysu.edu.cn} \maketitle

\begin{abstract}
In this paper, new enumerating functions for linear codes are defined,
including the triangle enumerating function and the tetrahedron enumerating function,
both of which can be computed using a trellis-based algorithm over polynomial rings.
The computational complexity is dominated by the complexity of the trellis.
In addition, we show that these new enumerating functions can be used to improve
existing performance bounds on the maximum likelihood decoding.
\end{abstract}

\section{Introduction}\label{introduction}

The weight enumerating function~(WEF)~\cite{MacWilliams77} is a
figure of merit of a linear code, which plays a fundamental rule in
the performance analysis of the maximum likelihood~(ML) decoding
algorithm. The conventional union bound, which involves only
pair-wise error probabilities, is simple but loose and even diverges
in the low signal-to-noise ratio~(SNR) region.
One general methodology to improve the conventional union bound, as
shown in~\cite{Sason06}, is invoking the Gallager's first bounding
technique~(GFBT)

\begin{equation}
  {\rm Pr} \{E\} \leq {\rm Pr} \{E,\underline{y}\in\mathcal{R}\} + {\rm Pr}
\{\underline{y}\notin \mathcal{R}\}\label{Gallager_first},
\end{equation}
%
%
where $E$ denotes the error event, $\underline{y}$ denotes the
received signal vector, and $\mathcal{R}$ denotes an arbitrary
region around the transmitted signal vector which is usually
interpreted as the ``good region". Most existing upper bounds within
this framework,
say,~\cite{Berlekamp80,Kasami92,Sphere94,TSB94,Zangl01,Divsalar03,Yousefi04,Mehrabian06},
first choose the region $\mathcal{R}$ such that the second term
of~(\ref{Gallager_first}) can be easily evaluated and then derive
upper bounds on the first term of~(\ref{Gallager_first}) by using
(conditional) pair-wise error probabilities and the whole~(or
truncated) WEF of the code.

Yousefi and Khandani~\cite{Yousefi04} derived an improved upper bound by using a Bonferroni-type
inequality of the second degree instead of the union bound. Since the resulting upper bound cannot be
calculated in terms of the distance spectrum of the code, the original codebook is enlarged by all $n$-tuples of Hamming weight $w$,
resulting in a bound that is solely dependent on the distance spectrum but becomes looser.
Very similarly, Ma~{\em et al}~\cite{Ma11}
proposed using triplet-wise error probabilities instead of pair-wise
error probabilities to improve the union bound. To make the proposed bound computable in terms of the distance spectrum of the code, an upper bound on the triplet-wise error probability is derived in~\cite[Lemma 4]{Ma11}. It has been shown that the union bound based on the triplet-wise error probability is tighter than the conventional union bound~\cite[Theorem 1]{Ma11}.

This paper is concerned with further tightening the union bound by
alleviating the repeated accumulations caused by the use of the
pair-wise error probabilities. The basic approach is to explore more
detailed geometrical structure~(beyond the distance spectrum) of the
code when upper bounding the error probabilities. The main results
as well as the structure of this paper are summarized as follows.
\begin{enumerate}
\item In Sec.~\ref{sec2}, we define two new enumerating functions for linear codes, the triangle spectrum and the tetrahedron spectrum, both of which can be calculated by a trellis-based algorithm.
\item In Sec.~\ref{sec3}, we derive improved union bounds based the triangle spectrum and the tetrahedron spectrum of binary linear
codes. A toy example is given to show that the improvement is
possible in the low-SNR region, as expected. The proposed union
bound may be combined with other upper bounding techniques based on
GFBT, potentially resulting in tighter upper bounds.
  \item Sec.~\ref{conclusion} concludes this paper.
\end{enumerate}

\section{New Spectra of Linear Block Codes}\label{sec2}

Let $\mathbb{F}_q$ be the finite field of size $q$. Let
$\mathbb{F}_q^n$ denote the $n$-dimensional vector space consisting
of $n$-tuples over $\mathbb{F}_q$. Given $\underline v
\stackrel{\Delta}{=} (v_0, v_1, \cdots, v_{n-1}) \in
\mathbb{F}_q^n$, the number of non-zero component of $\underline v$,
denoted by $W_H(\underline v)$, is called the Hamming weight of
$\underline v$. The Hamming distance between two vectors $\underline
v$ and $\underline w$ is defined as $W_H(\underline w - \underline
v)$. A linear code $\mathcal{C}_q[n, k]$ is defined as a
$k$-dimensional linear subspace of $\mathbb{F}_q^n$. A vector in
$\mathcal{C}_q[n, k]$ is called a codeword. There are $q^k$ in total
codewords in $\mathcal{C}_q[n ,k]$, which are simply indexed by
$\underline c^{(i)}$, $0\leq i\leq q^k - 1$. Specifically, we use
$\underline c^{(0)}$ to represent the all-zero codeword.

\subsection{Weight Enumerating Function}

\begin{definition}\label{definition_weight_spectrum}
The {\em weight enumerating function~(WEF)} of $\mathcal{C}_q[n, k]$
is defined as~\cite{MacWilliams77}
\begin{equation}\label{WEF}
    A(X) \stackrel{\Delta}{=} \sum_{i}A_{i}X^{i},
\end{equation}
where $X$ is a dummy variable and $A_{i}$ denotes the number
of codewords having Hamming weight $i$.
\end{definition}

The sequence $\{A_i, 0\leq i\leq n\}$ is also called {\em weight
spectrum} of the code, which exhibits how many codewords that are $i$
positions far away from the reference codeword $\underline c^{(0)}$.
By linearity, we know that the weight spectrum is irrelevant to the
reference codeword. Clearly, we have
\begin{equation}\label{WS}
   \sum_{1\leq i \leq n} A_i = q^k-1.
\end{equation}
For a binary code with the all-one codeword, we further have $A_i = A_{n-i}$ for $0\leq i \leq n$.

\subsection{Triangle Enumerating Function}
\begin{definition}\label{definition_triangle_spectrum}
Let $\underline c^{(0)}$ be the all-zero codeword and $\underline c^{(1)}$ be an arbitrarily given non-zero codeword. The {\em triangle enumerating function~(TrEF)} of $\mathcal{C}_q[n, k]$ is defined as
\begin{equation}\label{TrEF}
    B(\underline c^{(1)}; X,Y) \stackrel{\Delta}{=} \sum_{i,j}B_{i,j}(\underline c^{(1)})X^{i}Y^{j},
\end{equation}
where $X,Y$ are two dummy variables and $B_{i,j}(\underline c^{(1)})$ denotes the number
codewords $\underline c$ satisfying $W_H(\underline c - \underline c^{(0)}) = i$ and
$W_H(\underline c - \underline c^{(1)}) = j$.
\end{definition}

Generally, the TrEF depends on the choice of the reference codeword
$\underline c^{(1)}$. When the context is clear, we may drop
the reference codeword from the notation. The sequence $\{B_{i,j}, 0\leq
i, j \leq n\}$ is also called the {\em triangle spectrum} of the
code. Clearly, we have
\begin{equation}\label{TrS}
    \sum_{1\leq i, j\leq n} B_{i, j} = q^k-2.
\end{equation}

For binary codes with the all-one codeword, we have the following proposition.
\begin{proposition}
Suppose that $\mathcal{C}_2[n, k]$ has the WEF $A(X)$ such that $A_n = 1$. Let $\underline c^{(1)}$ be the codeword of weight $n$. Then
\begin{equation}\label{TrEFforSpecialBinary}
    B_{i, j} = \left\{\begin{array}{rl}
                                           A_i, & i+j = n \\
                                           0, & i+j \neq n
                                         \end{array}\right..
\end{equation}
\end{proposition}

\begin{IEEEproof}
It can be proved by noticing that $W_H(\underline c -\underline c^{(0)}) = i$ if and only if $W_H(\underline c-\underline c^{(1)}) = n-i$.
\end{IEEEproof}

\subsection{Tetrahedron Enumerating Function}
\begin{definition}\label{definition_tetrahedron_spectrum}
Let $\underline c^{(0)}$ be the all-zero codeword. Let $\underline c^{(1)}$ and $\underline c^{(2)}$ be
two arbitrarily given codewords. The {\em tetrahedron enumerating function~(TeEF)} of the code $\mathcal{C}_q[n, k]$ is defined as
\begin{equation}\label{TeEF}
    C(\underline c^{(1)}, \underline c^{(2)}; X,Y,Z) \stackrel{\Delta}{=} \sum_{i,j,h}C_{i,j,h}(\underline c^{(1)}, \underline c^{(2)})X^{i}Y^{j}Z^{h},
\end{equation}
where $X,Y,Z$ are three dummy variables and $C_{i,j,h}(\underline c^{(1)}, \underline c^{(2)})$ denotes the
number of codewords $\underline c$ satisfying $W_H(\underline c - \underline
c^{(0)}) = i$, $W_H(\underline c - \underline c^{(1)}) = j$
and $W_H(\underline c - \underline c^{(2)}) = h$.
\end{definition}

Generally, the TeEF depends on the choice of the reference codewords
$\underline c^{(1)}$ and $\underline c^{(2)}$. When the context is
clear, we may drop the reference codewords from the notation. The
sequence $\{C_{i,j,h}, 0 \leq i, j, h \leq n\}$ is also called the
{\em tetrahedron spectrum} of the code. Clearly, we have
\begin{equation}\label{TeS}
    \sum_{1\leq i, j, h\leq n} C_{i, j, h} = q^k-3.
\end{equation}

\subsection{An Example}

We take the Hamming code $\mathcal{C}_2[7, 4]$ as an example to illustrate the introduced enumerating functions.

The WEF is $$A(X) = 1 + 7X^3 + 7X^4 + X^7.$$

Since the TrEF depends on the choice of the reference codeword $\underline c^{(1)}$, we distinguish following three cases.

\begin{enumerate}
  \item[]{\em Case 1:} If $W_H(\underline c^{(1)}) = 7$,
  $$B(X,Y) = Y^7 + 7X^3Y^4 + 7X^4Y^3 + X^7.$$
  \item[]{\em Case 2:} If $W_H(\underline c^{(1)}) = 4$,
  $$B(X,Y) = Y^4 + X^4 + 6X^3Y^3 + X^3Y^7 + 6X^4Y^4 + X^7Y^3.$$
  \item[]{\em Case 3:} If $W_H(\underline c^{(1)}) = 3$,
  $$B(X,Y) = Y^3 + X^3 + 6X^3Y^4 + 6X^4Y^3 + X^4Y^7 + X^7Y^4.$$
\end{enumerate}

Similarly, the TeEF also depends on the choices of the reference codewords $\underline c^{(1)}$ and $\underline c^{(2)}$. We have

\begin{enumerate}
  \item[]{\em Case 1:} If $W_H(\underline c^{(1)}) = 3$ and $W_H(\underline c^{(2)}) = 3$,
   \begin{equation*}
    \begin{array}{l}
      C(X, Y, Z) =  Y^3Z^3 + X^3Y^4 + X^3Z^4 + \\
      5X^3Y^4Z^4 + 5X^4Y^3Z^3 + X^4Y^3Z^7 + X^4Y^7Z^3 + X^7Y^4Z^4.
    \end{array}
   \end{equation*}

  \item[]{\em Case 2:} If $W_H(\underline c^{(1)}) = 3$, $W_H(\underline c^{(2)}) = 4$, and $W_H(\underline c^{(2)}-\underline c^{(1)}) = 3$,
  \begin{equation*}
    \begin{array}{l}
      C(X, Y, Z) =  Y^3Z^4 + X^3Z^3 + X^4Y^3 + \\
      5X^3Y^4Z^3 + X^3Y^4Z^7 + 5X^4Y^3Z^4 + X^4Y^7Z^4 + X^7Y^4Z^3.
    \end{array}
   \end{equation*}

  \item[]{\em Case 2':} If $W_H(\underline c^{(1)}) = 3$, $W_H(\underline c^{(2)}) = 4$, and $W_H(\underline c^{(2)}-\underline c^{(1)}) = 7$,
  \begin{equation*}
    \begin{array}{l}
      C(X, Y, Z) =  Y^3Z^4 + X^3Z^7 + X^4Y^7 + \\
      6X^3Y^4Z^3 + 6X^4Y^3Z^4 + X^7Y^4Z^3.
    \end{array}
   \end{equation*}

  \item[]{\em Case 3:} If $W_H(\underline c^{(1)}) = 3$ and $W_H(\underline c^{(2)}) = 7$,
  \begin{equation*}
    \begin{array}{l}
      C(X, Y, Z) =  Y^3Z^7 + X^3Z^4 + X^7Y^4 + \\
      6X^3Y^4Z^4 + 6X^4Y^3Z^3 + X^4Y^7Z^3.
    \end{array}
   \end{equation*}

  \item[]{\em Case 4:} If $W_H(\underline c^{(1)}) = 4$ and $W_H(\underline c^{(2)}) = 4$,
  \begin{equation*}
    \begin{array}{l}
      C(X, Y, Z) =  Y^4Z^4 + X^4Z^4 + X^4Y^4 + \\
      5X^3Y^3Z^3 + X^3Y^3Z^7 + X^3Y^7Z^3 + 5X^4Y^4Z^4 + X^7Y^3Z^3.
    \end{array}
   \end{equation*}
   \item[]{\em Case 5:} If $W_H(\underline c^{(1)}) = 4$ and $W_H(\underline c^{(2)}) = 7$,
  \begin{equation*}
    \begin{array}{l}
      C(X, Y, Z) =  Y^4Z^7 + X^4Z^3 + X^7Y^3 + \\
      6X^3Y^3Z^4 + X^3Y^7Z^4 + 6X^4Y^4Z^3.
    \end{array}
   \end{equation*}

\end{enumerate}


\subsection{Computing the Enumerating Functions Over a Trellis}

It is well-known that any linear block code can be represented by a
trellis~\cite{McEliece96}~\cite{Vardy98}.
Generally, a trellis that represents $\mathcal{C}_q[n, k]$ can have $N$ stages. The trellis section at stage $t$~($0\leq t \leq N-1$), denoted by  $\mathcal{B}_t$, is a subset of $\mathcal{S}_{t} \times \mathbb{F}_q^{n_t} \times \mathcal{S}_{t+1}$, where $\mathcal{S}_t$ is the state space at time $t$. A branch $b \in \mathcal{B}_{t}$ is denoted by $b\stackrel{\Delta}{=} (\sigma^-(b), \ell(b), \sigma^+(b))$, starting from a state $\sigma^-(b)\in \mathcal{S}_{t}$, taking a label $\ell(b) \in \mathbb{F}_q^{n_t}$, and ending into a state $\sigma^+(b)\in \mathcal{S}_{t+1}$. A path through a trellis is a sequence of branches ${\underline b} = (b_0, b_1, \cdots, b_{N-1})$ satisfying that $b_t \in \mathcal{B}_t$ and $\sigma^-(b_{t+1}) = \sigma^+(b_{t})$.  A codeword is then represented by a path in the sense that $\underline c = (\ell(b_0), \ell(b_1), \cdots, \ell(b_{N-1}))$. Naturally, $\sum_{0\leq t \leq N-1}n_t = n$ and the number of paths is $q^k$. Without loss of generality, we set $\mathcal{S}_0 = \mathcal{S}_N = \{0\}$.

\begin{proposition}\label{Proposition_trellis}
Given a trellis representation of $\mathcal{C}_q[n, k]$. Let $\underline c^{(0)}$~(the all-zero codeword),  $\underline c^{(1)}$ and  $\underline c^{(2)}$ be three reference codewords. The corresponding pathes are denoted by ${\underline b}^{(0)}  = (b_0^{(0)}, b_1^{(0)}, \cdots, b_{N-1}^{(0)})$, ${\underline b}^{(1)}  = (b_0^{(1)}, b_1^{(1)}, \cdots, b_{N-1}^{(1)})$ and ${\underline b}^{(2)}  = (b_0^{(2)}, b_1^{(2)}, \cdots, b_{N-1}^{(2)})$, respectively. Then the enumerating function~(WEF, TrEF or TeEF) is equal to $\alpha_N(0)$, as calculated recursively by
the following trellis-based algorithm over a properly defined polynomial ring.
\begin{itemize}
  \item Initially, set $\alpha_0(0) = 1$.
  \item For $t = 0, 1, \cdots, N-1$,
        \begin{eqnarray}
            \alpha_{t+1}(s) &=& \sum_{b\in\mathcal{B}_t, \sigma^+(b) =
            s}\alpha_{t}(\sigma^-(b))\gamma_t(b)
        \end{eqnarray}
        for each state $s \in \mathcal{S}_{t+1}$, where $\gamma_t(b)$ is specified as follows.
        \begin{enumerate}
        \item[]{\em Case 1:}  For computing WEF, $\gamma_t(b) \stackrel{\Delta}{=} X^i$, where $i = W_H(\ell(b))$.
        \item[]{\em Case 2:} For computing TrEF, $\gamma_t(b) \stackrel{\Delta}{=} X^iY^j$, where $i = W_H(\ell(b))$ and $j = W_H(\ell(b) - \ell(b_t^{(1)}))$.
        \item[]{\em Case 3:} For computing TeEF, $\gamma_t(b) \stackrel{\Delta}{=} X^iY^jZ^h$, where $i = W_H(\ell(b))$, $j = W_H(\ell(b) - \ell(b_t^{(1)}))$ and $h = W_H(\ell(b) - \ell(b_t^{(2)}))$.
       \end{enumerate}
  \end{itemize}
\end{proposition}

\begin{IEEEproof}
The algorithm is similar to the trellis algorithm over polynomial
rings for computing the weight enumerators of
paths~\cite{McEliece96}.
\end{IEEEproof}

{\bf Remark.} It can be seen that the computational complexity of the algorithm given in Proposition~\ref{Proposition_trellis} is dominated by the complexity of the trellis~\cite{McEliece96}. From this algorithm, we also know that $B(\underline c^{(1)}; X, Y) = C(\underline c^{(1)}, \underline c^{(2)}; X, Y, Z = 1)$ and $A(X) = B(\underline c^{(1)}; X, Y = 1)$.

\section{Improved Union Bounds for Binary Linear Codes Based on Geometrical Spectra}\label{sec3}
In this section, we focus on tightening the conventional union bound based on pair-wise error probabilities by exploring
further the geometrical structure of codes.

\subsection{Geometrical Properties of Binary Codes}
Let $\mathbb{F}_2 = \{0, 1\}$ and $\mathcal{A}_2 = \{-1, +1\}$ be
the binary field and the bipolar signal set, respectively. Suppose
that a codeword $\underline{c}=(c_0, c_1, \cdots, c_{n-1}) \in
\mathcal{C}_2[n,k]$ is modulated by binary phase shift
keying~(BPSK), resulting in a bipolar signal vector $\underline s
\in \mathcal{A}_2^n$ with $s_t = 1 - 2c_t$ for $0\leq t \leq n-1$.
We will not distinguish between a binary codeword $\underline c$ and its bipolar image in the following, except when we need to emphasize the difference between the Hamming space $\mathbb{F}_2^n$ and the Euclidean space $\mathbb{R}^n \supset \mathcal{A}_2^n$. The Euclidean distance between two codewords $\underline s^{(1)}$ and $\underline s^{(2)}$ is related to their Hamming distance by
$\| \underline s^{(2)} - \underline s^{(1)} \| = 2\sqrt{W_H(\underline c^{(2)} - \underline c^{(1)})}$. All codewords are distributed on the surface of  an $n$-dimensional sphere centered at the origin with radius $\sqrt{n}$. This property is referred to as the {\em sphericity} of the bipolar code.

Assume that a codeword $\underline s$ is transmitted over an AWGN channel, resulting in a received vector $\underline{y} = {\underline s} + {\underline z}$, where $\underline z$ is a sample from a white
Gaussian noise process with zero mean and double-sided power
spectral density $\sigma^2$. The ML decoding is equivalent to finding a bipolar codeword $\underline s$ that is the closest to $\underline y$. Since the decoding metric is the Euclidean distance, the geometrical structure of the code in $\mathbb{R}^n$ is supposed to be critical to analyze the ML decoding performance. However, to the best knowledge of ours, with the exception of the distance spectrum and the sphericity of the code, other figures of merits of the code were rarely employed to upper bound the ML decoding error probability. To reveal more information about the geometrical structure of the code, we have the following two propositions, where Proposition~\ref{Pro_Three_codewords} was originally mentioned in~\cite{Agrell98} without proofs.
\begin{proposition}\label{Pro_Three_codewords}
Any three codewords form a non-obtuse triangle. Furthermore, if
some three codewords form a right angle, there must exist a fourth codeword
completing the rectangle.
\end{proposition}

\begin{IEEEproof}
For a detailed proof of the first part, see~\cite{Ma11}.

To prove the second part, we may assume by linearity that
$\underline s^{(0)}$, $\underline s^{(1)}$ and $\underline s^{(2)}$
form a right angle, that is,
$\overrightarrow{\underline{s}^{(0)}\underline{s}^{(1)}}$ is
orthogonal to
$\overrightarrow{\underline{s}^{(0)}\underline{s}^{(2)}}$. Noting
that this holds if and only if $W_H(\underline c^{(1)} + \underline
c^{(2)}) = W_H(\underline c^{(1)}) + W_H(\underline c^{(2)})$,
implying that the two codewords $\underline c^{(1)}$ and $\underline
c^{(2)}$ are not ``overlapped"~(no common non-zero positions). Hence
the binary addition $\underline c^{(1)} + \underline c^{(2)}$ can be
treated as a real addition. Define the codeword $\underline c^{(3)}
= \underline c^{(1)} + \underline c^{(2)}$. We can verify that
$$\underline s^{(3)} - \underline s^{(0)} = \underline s^{(1)} - \underline s^{(0)} + \underline s^{(2)} - \underline s^{(0)},$$
which means that $\overrightarrow{\underline s^{(0)}\underline s^{(3)}}$ falls inside the plane determined by
$\overrightarrow{\underline s^{(0)}\underline s^{(1)}}$ and $\overrightarrow{\underline s^{(0)}\underline s^{(2)}}$ and hence $\underline s^{(0)}$, $\underline s^{(1)}$, $\underline s^{(2)}$ and $\underline s^{(3)}$ must form a rectangle. Otherwise, some three of them would form an obtuse triangle. \end{IEEEproof}

\begin{proposition}
Any four codewords form either a tetrahedron or a rectangle.
\end{proposition}

\begin{IEEEproof}
From Proposition~\ref{Pro_Three_codewords}, any three
codewords form a non-obtuse triangle, which determines a two-dimensional plane. If the fourth codeword falls inside the same plane, the four codewords must form a rectangle; otherwise, some three of them would form an obtuse triangle. If the fourth
codeword falls outside that plane, then the four codewords form
a tetrahedron in a three-dimensional space.
\end{IEEEproof}

\begin{figure}
\centering
  \includegraphics[width=11cm]{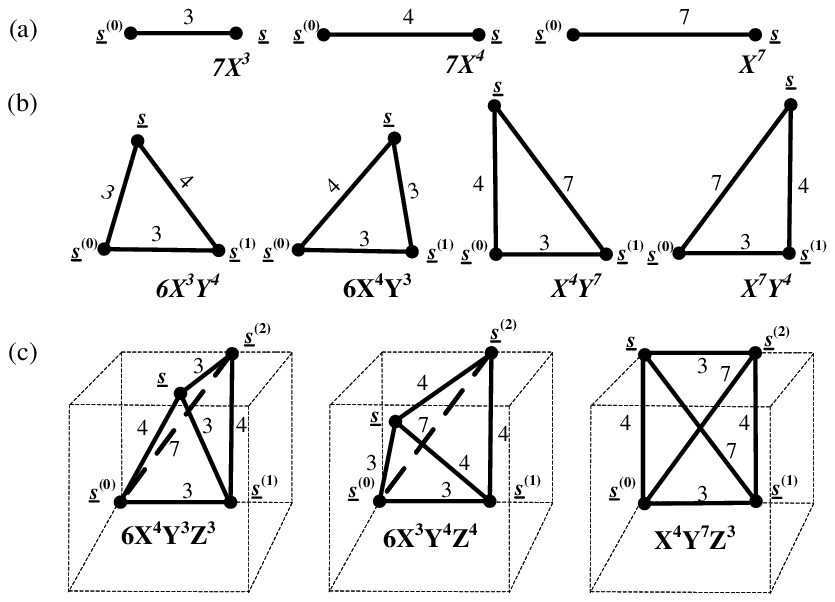}\\
  \caption{Geometrical spectra of the Hamming code $\mathcal{C}_2[7, 4]$, where $\underline s^{(i)}$, $i = 0, 1, 2$, are taken as the reference codewords and each edge is labeled by the Hamming distance. (a)~The weight spectrum. (b)~A triangle spectrum. (c)~A tetrahedron spectrum.}\label{Fig_7_4_3_EF}
\end{figure}

With BPSK signalling, we also refer WEF, TrEF and TeEF  to as {\em geometrical spectra} of a code. Fig.~\ref{Fig_7_4_3_EF} shows the geometrical spectra of the Hamming code $\mathcal{C}_2[7, 4]$.

\subsection{Improved Union Bounds Based on Geometrical Spectra}
Assume that ${\underline s}^{(0)}$ is transmitted. For a codeword $\underline s$, let
$$\{\underline s^{(0)} \rightarrow \underline s\} \stackrel{\Delta}{=} \{\underline y: \|\underline y - \underline s\| \leq \|\underline y - \underline s^{(0)} \|\},$$
which is the event that $\underline s$ is nearer than $\underline{s}^{(0)}$ to $\underline{y}$. We use $\{\underline s^{(0)} \nrightarrow \underline s\}$ to denote the complementary event.

To derive the upper bounds on the decoding error probability ${\rm Pr}\{E\}$, we take two arbitrary but fixed codewords ${\underline s^{(1)}}$ and ${\underline s^{(2)}}$ as reference codewords. Let $d_1 = W_H(\underline c^{(1)})$, $d_2 = W_H(\underline c^{(2)})$ and $d_{1, 2} = W_H(\underline c^{(1)} - \underline c^{(2)})$. For a codeword $\underline c$, let $i = W_H(\underline c - \underline c^{(0)})$, $j = W_H(\underline c - \underline c^{(1)})$ and $h = W_H(\underline c - \underline c^{(2)})$.
It is well-known that the {\em pair-wise error probability~(PEP)} $p_2(i) \stackrel{\Delta}{=}{\rm Pr}\{\underline s^{(0)} \rightarrow \underline s\}$ is given by $Q(\sqrt{i}/\sigma)$ and depends solely on the Hamming weight. Going a step further, we can verify that the {\em triplet-wise error probability~(TrEP)}, defined by
$$p_3(i, j)\stackrel{\Delta}{=}{\rm Pr}\left\{(\underline s^{(0)} \rightarrow \underline s^{(1)}) \bigcup (\underline s^{(0)} \rightarrow \underline s)\right\},$$
depends solely on the triangle formed by the three codewords. Similarly, the {\em quadruple-wise error probability~(QuEP)}, defined by
$$p_4(i, j, h)\stackrel{\Delta}{=}{\rm Pr}\left\{(\underline s^{(0)} \rightarrow \underline s^{(1)})\bigcup (\underline s^{(0)} \rightarrow \underline s^{(2)}) \bigcup (\underline s^{(0)} \rightarrow \underline s)\right\},$$
depends solely on the tetrahedron~(or rectangle) formed by the four codewords. For these reasons, we have dropped the codeword $\underline s$ from the notation and simply denoted these probabilities by $p_2(i)$, $p_3(i, j)$ and $p_4(i, j, h)$ as shown above.

\begin{figure}
\centering
  \includegraphics[width=11.5cm]{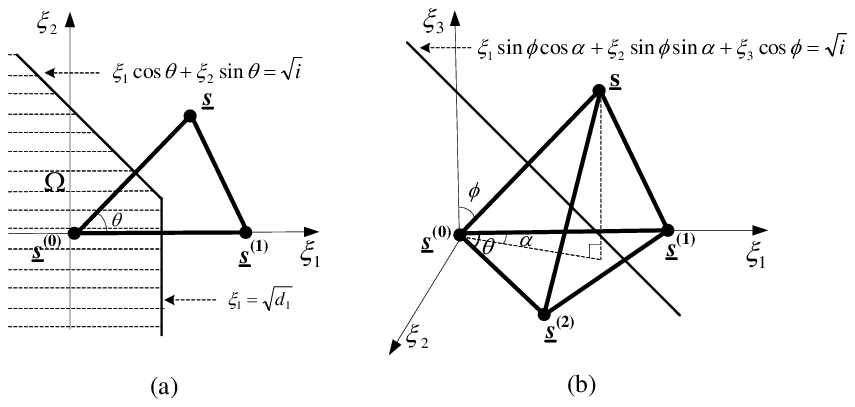}\\
  \caption{Geometrical interpretation of error probabilities. (a)~Triplet-wise error probability. (b)~Quadruple-wise error probability.}\label{Fig_TrEP_QuEP}
\end{figure}

To compute the introduced error probabilities conveniently, we may use a new coordinate system by choosing $\underline s^{(0)}$ as the origin $O$ and taking $\overrightarrow{\underline s^{(0)}\underline s^{(1)}}$ as an axis, denoted by $\xi_1$-coordinate. We further choose $\xi_2$-coordinate such that $\underline s^{(2)}$ falls into the first quadrant of the plane $\xi_1O\xi_2$. Similarly, we choose $\xi_3$-coordinate such that the fourth codeword $\underline s$ falls in the first octant, as shown in Fig.~\ref{Fig_TrEP_QuEP}. Note that such an arrangement does not lose any generality. Let $Z_{\xi_1}$, $Z_{\xi_2}$, and $Z_{\xi_3}$ be the three components obtained by projecting the noise $\underline Z$ onto the three axes, respectively. Specifically, $Z_{\xi_1}$ is the inner product $\langle
{\underline Z}, \frac{\underline{s}^{(1)}-\underline{s}^{(0)}}{\|\underline{s}^{(1)}-\underline{s}^{(0)}\|}\rangle$.
These three component are independent and identically distributed as a Gaussian random variable with a probability density function $f(x) = \frac{1}{\sqrt{2\pi}\sigma}\exp \{-\frac{x^2}{2 \sigma^2}\}$. We have the following lemmas.

\begin{lemma}\label{Lemma_TrEP}
The TrEP can be calculated as
\begin{equation}\label{triplet-wise}
    p_3(i, j) = 1 - \int\int_{\Omega}f(\xi_1) f(\xi_2)~{\rm
d}\xi_1~{\rm d}\xi_2,
\end{equation}
where $\Omega = \{\xi_1 < \sqrt{d_1},~\xi_1 \cos\theta + \xi_2\sin\theta < \sqrt{i} \}$ and $\cos \theta = (d_1 + i -j) / (2\sqrt{d_1i})$.
\end{lemma}

\begin{IEEEproof} It can be proved by verifying that, given the three codewords, $\Omega$ is exactly the Voronoi region of $\underline s^{(0)}$. See Fig.~\ref{Fig_TrEP_QuEP}~(a) for a reference.
%
%
\end{IEEEproof}

%
\begin{lemma}\label{Lemma_QuEP}
The QuEP can be calculated as
\begin{equation}\label{quadruple-wise}
    p_4(i, j, h) = 1 - \int\!\!\int\!\!\int_\Omega\!\! f(\xi_1)f(\xi_2)f(\xi_3)~{\rm d}\xi_1~{\rm
d}\xi_2~{\rm d}\xi_3.
\end{equation}
The integration domain
$$\Omega = \left\{
\begin{array}{l}
  \xi_1 < \sqrt{d_1}, ~~\xi_1 \cos \theta + \xi_2 \sin \theta < \sqrt{d_2},\\
\xi_1 \sin \phi \cos \alpha + \xi_2 \sin \phi \sin \alpha + \xi_3 \cos\phi < \sqrt{i}
\end{array}\right\}
$$
can be determined by computing the azimuth angle $\theta$ of $\underline s^{(2)}$,  the azimuth angle
$\alpha$ of $\underline s$ and the colatitude angle $\phi$ of $\underline s$. See Fig.~\ref{Fig_TrEP_QuEP}~(b) for a reference.
\end{lemma}

\begin{IEEEproof}
It can be proved by verifying that, given the four codewords, $\Omega$ is exactly the Voronoi region of $\underline s^{(0)}$.
\end{IEEEproof}

{\bf Remark.} Note that the angles appeared in
Lemma~\ref{Lemma_QuEP} are computable given the edge lengths of the
tetrahedron. For example, $\theta$ can be computed by the law of
cosines: $\cos \theta = (d_1+d_2-d_{1,2})/(2\sqrt{d_1d_2})$. And the
expressions for $\alpha$ and $\phi$ can be obtained by

$$\alpha= \arctan\left( \frac{\sqrt{(i+d_1-j)^2\cos
^2\theta+\frac{d_1(i+d_2-h)^2}{d_2}-\frac{2\sqrt{d_1}\cos\theta(i+d_1-j)(i+d_2-h)}{\sqrt{d_2}}}}{(i+d_1-j)\sin\theta}
\right)$$ and

$$\phi = \arcsin\left(
\frac{\sqrt{4i\sin^2\theta-\frac{(i+d_1-j)^2}{d_1}
-\frac{(i+d_2-h)^2}{d_2} +
\frac{2\cos\theta(i+d_1-j)(i+d_2-h)}{\sqrt{d_1d_2}} }}{2\sqrt
{i}\sin\theta} \right),$$ respectively.

Also note that Lemma~\ref{Lemma_QuEP} is still valid in the case
when the four codewords form a rectangle. It is worth pointing out
the both TrEP and QuEP can be transformed into repeated integrals
easily.

\begin{theorem}\label{Theorem_union_triangle}
Let $\underline c^{(1)}$ be any fixed reference codeword with $W_H(\underline
c^{(1)}) = d_1 \geq 1$. Assume that the corresponding triangle spectrum $\{B_{i, j}\}$ is available. The ML decoding error can be upper bounded by
$${\rm Pr}\left\{E\right\} \leq   - (2^k-3) Q(\sqrt{d_1}/\sigma) + \sum_{1\leq i, j \leq n}B_{i, j} p_3(i, j),$$
where $p_3(i, j)$ are given by~(\ref{triplet-wise}).
\end{theorem}

\begin{IEEEproof}
From the second-order Bonferroni-type inequality, we have
$$
\begin{array}{l}
  {\rm Pr}\{E\} = {\rm Pr}\left\{\bigcup_{\underline s \neq \underline s^{(0)}}(\underline s^{(0)}\rightarrow \underline s)\right\}\\
  \leq {\rm Pr}\{\underline s^{(0)}\rightarrow \underline s^{(1)}\} +
 \sum'{\rm Pr}\left\{\underline s^{(0)}\nrightarrow \underline s^{(1)}, \underline s^{(0)}\rightarrow \underline s\right\}\\
  = -(2^k - 3) {\rm Pr}\{\underline s^{(0)}\rightarrow \underline s^{(1)}\}
  + \sum'{\rm Pr}\left\{(\underline s^{(0)}\rightarrow \underline s^{(1)}) \bigcup (\underline s^{(0)}\rightarrow \underline s)\right\},
\end{array}
$$
where the summation $\sum'$ is over all $\{\underline s: \underline s \neq \underline s^{(0)}, \underline s \neq \underline s^{(1)}\}$. This completes the proof by noting that the TrEP depends only on the types of the triangles.
\end{IEEEproof}

\begin{theorem}\label{Theorem_union_tetrahedron}
Let $\underline c^{(1)}$ and $\underline c^{(2)}$ be any two fixed reference codewords with $W_H(\underline
c^{(1)}) = d_1 \geq 1$, $W_H(\underline
c^{(2)}) = d_2 \geq 1$ and $W_H(\underline
c^{(2)} - \underline
c^{(1)}) = d_{1, 2} \geq 1$. Assume that the corresponding tetrahedron spectrum $\{C_{i, j, h}\}$ is available. The ML decoding error can be upper bounded by
$${\rm Pr}\left\{E\right\} \leq   - (2^k-4) p_3(d_2, d_{1, 2}) + \sum_{1\leq i, j, h \leq n}C_{i, j, h} p_4(i, j, h),$$
where $p_3(d_2, d_{1,2})$ and $p_4(i, j, h)$ are given by~(\ref{triplet-wise}) and~(\ref{quadruple-wise}), respectively.
\end{theorem}

\begin{IEEEproof}
From the third-order Bonferroni-type inequality, we have
$$
\begin{array}{l}
  {\rm Pr}\{E\} = {\rm Pr}\left\{\bigcup_{\underline s \neq \underline s^{(0)}}(\underline s^{(0)}\rightarrow \underline s)\right\} \\
  \leq {\rm Pr}\{(\underline s^{(0)}\rightarrow \underline s^{(1)}) \bigcup (\underline s^{(0)}\rightarrow \underline s^{(2)})\} + \\ \sum\limits_{\underline s \neq \underline s^{(i)}, i = 0, 1, 2}{\rm Pr}\left\{\underline s^{(0)}\nrightarrow \underline s^{(1)}, \underline s^{(0)}\nrightarrow \underline s^{(2)}, \underline s^{(0)}\rightarrow \underline s\right\}\\
  = -(2^k - 4) {\rm Pr}\{(\underline s^{(0)}\rightarrow \underline s^{(1)}) \bigcup (\underline s^{(0)}\rightarrow \underline s^{(2)})\} \\
  + \sum\limits_{\underline s \neq \underline s^{(i)}, i = 0, 1, 2}{\rm Pr}\left\{(\underline s^{(0)}\rightarrow \underline s^{(1)}) \bigcup (\underline s^{(0)}\rightarrow \underline s^{(2)}) \bigcup (\underline s^{(0)}\rightarrow \underline s)\right\},
\end{array}
$$
completing the proof.
\end{IEEEproof}

\begin{figure}
\centering
  \includegraphics[width=9cm]{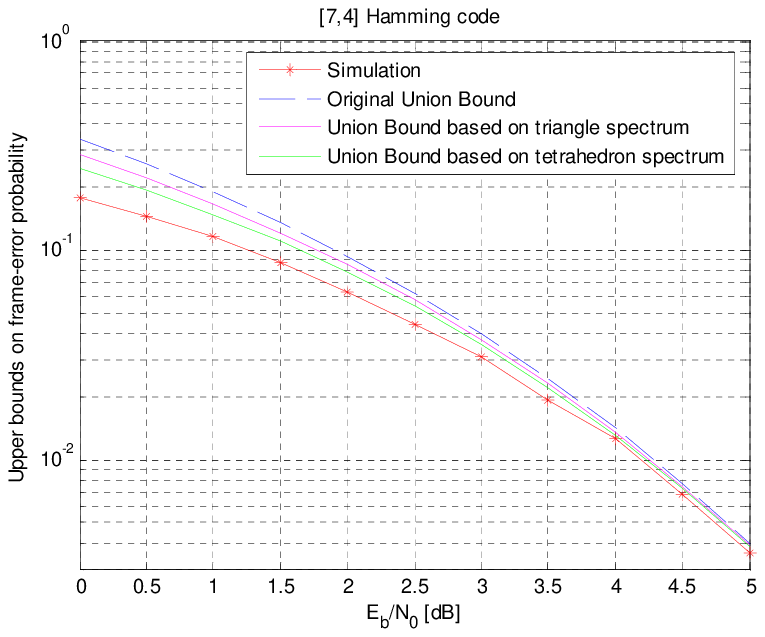}\\
  \caption{Comparison between the upper bounds on the frame-error probability
under ML decoding of [7, 4] Hamming code. The compared bounds are
the original union bound, the union bound based on {\em triangle
spectrum} and the union bound based on {\em tetrahedron spectrum},
which are also compared with the ML simulation
results.}\label{Fig_hamming_code}
\end{figure}

\subsection{Numerical Results}
From the proofs of Theorems~\ref{Theorem_union_triangle}~and~\ref{Theorem_union_tetrahedron}, we know that the proposed bounds compute the higher-order Bonferroni-type inequalities. Hence the proposed bounds are tighter than the conventional union bound. To verify this numerically, we give an example.
Fig.~\ref{Fig_hamming_code} shows the comparisons between the
original union bound and the bounds  given in
Theorems~\ref{Theorem_union_triangle}~and~\ref{Theorem_union_tetrahedron} on the frame-error
probability of the Hamming code $\mathcal{C}_2[7, 4]$. Also shown
are the simulation results. The TrEF and TeEF we choose are
$B(X,Y) = Y^3 + X^3 +
6X^3Y^4 + 6X^4Y^3 + X^4Y^7 + X^7Y^4$ and $ C(X, Y, Z) =  Y^3Z^3 +
X^3Y^4 + X^3Z^4 + 5X^3Y^4Z^4 + 5X^4Y^3Z^3 + X^4Y^3Z^7 + X^4Y^7Z^3 +
X^7Y^4Z^4$, respectively.
We can see that the bounds using higher-order Bonferroni-type inequalities are tighter, as expected.

\section{Conclusions}\label{conclusion}
In this paper, we have presented the definition of TrEF and TeEF,
both of which can be computed using a trellis-based algorithm over
polynomial rings. We have also derived the upper bounds based on
{\em triangle spectrum} and {\em tetrahedron spectrum},
respectively, which can be used to improve the union bound by
alleviating the repeated accumulations caused by the use of the
pair-wise error probabilities.


\appendices

\ifCLASSOPTIONcaptionsoff

\newpage
\fi

 \small
\bibliographystyle{IEEEtran}
\bibliography{IEEEabrv,tzzt}

\end{document}